\documentclass[12pt]{article}
\usepackage{graphicx}
\begin{document}

version 14.2.2005 (accepted in Phys.~Rev.~C)
\begin{center}
{\Large Structure of the Breathing Mode of the Nucleon \\
from high energy p-p scattering} 

H.P. Morsch \\ 
Institut f\"ur Kernphysik, Forschungszentrum 
J\"ulich, D-52425 J\"ulich, Germany \\ P. Zupranski \\
Soltan Institute for Nuclear Studies, Pl-00681 Warsaw, Poland
\end{center}

\begin{abstract}
Spectra of p-p and $\pi$-p scattering at beam momenta between 6 and 30 GeV/c
have been reanalysed. These show strong excitation of N* resonances,
the strongest one corresponding to the ``scalar'' P$_{11}$ excitation  
(breathing mode) at $m_o=1400\pm 10$ MeV  with $\Gamma =200\pm 25$ MeV. 
The result of a strong scalar excitation is supported by a large 
longitudinal amplitude $S_{1/2}$ extracted from e-p scattering. 
From exclusive data on $p+p\rightarrow pp\ \pi^+\pi^-$ a large $2\pi$-N 
decay branch for the P$_{11}$ resonance of $B_{2\pi}=75\pm 20$~\% has been
extracted. 

The differential cross sections were described in a double-folding approach, 
assuming multi-gluon exchange as the dominant part of the effective 
interaction between the constituents of projectile and target. 
First, the parameters of the interaction were fitted to elastic scattering;
then with this interaction the inelastic cross sections were described in the 
distorted wave Born approximation. A good description of the data requires a
surface peaked transition density, quite different from that of a pure 
radial mode. In contrast, the electron scattering amplitude $S_{1/2}$ is 
quite well described by a breathing mode
transition density with radial node. This large difference between charge 
and matter transition densities suggests, that in p-p scattering 
the coupling to the multi-gluon field is much more important than the coupling
to the valence quarks. A multi-gluon (or sea-quark) transition density is
derived, which shows also breathing, indicating a rather complex multi-quark 
structure of N and N* including multi-glue (or $q^{2n}\bar q^{2n}$)
creation out of the g.s.~vacuum.
\end{abstract}

{\bf 1. Introduction}

The spectrum of baryon resonances is directly related to the 
dynamical structure of quantum chromodynamics, however, the origin of 
these resonances is so far not well understood. 
In particular, the lowest N* resonance, P$_{11}$ at a mass of about 
1400 MeV (Roper resonance), has been discussed controversially
in many different models, in the constituent quark model \cite{qqq,Gloz}, 
the Skyrmion model \cite{Skyrm}, bag models \cite{bag}, being of 
hybrid structure \cite{hybrid}, or generated by a strong $\sigma$-N coupling
\cite{sigma}. Recent calculations in lattice QCD \cite{latt} have shown,
that the mass of the lowest P$_{11}$ is reproduced in quenched
approximation, indicating that this resonance contains valence quark 
contributions. 

Experimentally, the lowest P$_{11}$ resonance exhibits  
a rather complex structure, which shows up differently in 
several reactions: whereas in $\pi$-N phase shift analyses 
\cite{phasepiN} this resonance is quite weak and has a large 
width in the order of 350 MeV, in $\alpha$-p scattering \cite{Mo92} a 
strong monopole excitation has been found in this mass region with a much 
smaller width of about 200 MeV. In inclusive electron scattering 
data~\cite{elec} a P$_{11}$ resonance has not been observed, however, in 
$\gamma$-induced processes it is rather small but clearly observed 
(see e.g.~\cite{gamma}). In an attempt to understand the
different results, a T-matrix description of $\alpha$-p, 
$\pi$-N, and $\gamma$-p has been performed \cite{MoZu}, which shows, 
that all data can be described consistently by assuming two 
structures in this resonance, a strong scalar monopole excitation (Saturne 
resonance) interpreted as breathing mode of the nucleon, 
and a second much wider structure, which can be interpreted as second order
$\Delta$ excitation (spin-isospin (M1) mode). 
Preliminary results from a more exclusive study of $\alpha$-p $\rightarrow 
\alpha'\ (p,\pi^+)\ x$ \cite{spes4pi} support these conclusions.

To study further N* resonances we have analysed old p-p and $\pi$-p scattering 
data \cite{Andp,Edel,And} taken at beam momenta above 5 GeV/c, in which a
pronounced resonance structure is observed. The analysis
of these spectra is discussed in sect.~2 with emphasis on the breathing mode 
excitation P$_{11}$ at 1400 MeV. The decay properties of this resonances
are discussed in sect.~3. Then an attempt is made in sects.~4 and 5 to
understand the breathing mode excitation in p-p and e-p, which necessitates 
the assumption of a strong multi-gluon field in the nucleon. From this 
phenomenological analysis new insight into the structure of scalar 
N* excitations is obtained.

{\bf 2. Analysis of p-p and $\pi$-p spectra}

In the p-p and $\pi$-p spectra of ref.~\cite{Andp,Edel,And} several resonance
structures have been observed, 
the $\Delta$(1232) and resonances at about 1400 MeV, 1520 MeV, 1680 MeV and 
2190 MeV. Remarkably, by far the strongest resonance (at small momentum
transfer) is found in the region of the lowest P$_{11}$.
It is important to mention, that the wealth of p and $\pi$ scattering
data in this energy region, taken at very different momentum transfers, showed 
evidence just for these few resonances (in contrast to low energy $\pi$-N, 
in which many more resonances show up). This indicates a {\bf high 
selectivity} in the excitation of N* resonances (with quantum
numbers J$^{\pi}$=L$\pm \frac{1}{2}^{(-)^L}$, where L is the angular momentum 
transfer). This can be explained by the fact, that multi-gluon exchange 
(of scalar character) is the predominant part of the interaction 
(see e.g.~\cite{MoSD}).

Missing mass spectra of p-p and $\pi$-p scattering are shown in figs.~1-4 
together with resonance fits given by solid lines. Apart from the P$_{11}$
the shapes were taken from a fit of $\pi$-N phase shift amplitudes 
(see ref.~\cite{MoZu,MoSD}), using modified Breit-Wigner forms.
Important is also a multi-pion background, which is described 
by 1$\pi$ and 2$\pi$ threshold functions of a form similar
to that used in ref.~\cite{MoZu}, together with a polynomial rise to larger 
masses:  
\begin{equation}
 back(m)=\sum_{n=1}^{2}\ [ c_n\ f_{n\pi} + d_n\ (m-m_{n\pi})^n] 
\end{equation} 
with $f_{n\pi}=[(\beta-1)/(\beta+1)]^n$ and $\beta=exp[\frac{(m-m_{n\pi})}
{(m_o-m_{n\pi})}*C_{n\pi}]$.
A value of $m_o$ of 1.35 GeV is needed, when strong $\Delta$ excitation 
is observed (at low beam momentum $\leq$10 GeV/c and small 
momentum transfer $\leq$0.05 (GeV/c)$^2$), otherwise $m_o$ 
is about 1.48 GeV (or 1.55 at -t$\geq$ 0.2 (GeV/c)$^2$). For $C_{1\pi}$ and 
$C_{2\pi}$ values of 15 and 4 were used.
We fitted the background in all p-p spectra from 6 to 20 GeV/c 
simultaneously with a smooth variation of the parameters $c_n$ and $d_n$ 
($c_1\sim$(5-10)$\cdot c_2$, $d_1\sim$1/3$\cdot c_2$, and $d_2$ very small). 
For a small momentum transfer of t=0.044 (GeV/c)$^2$ the threshold parameter 
$c_1$ falls off towards higher beam momenta by a factor of two, whereas $d_1$ 
decreases by about a factor of four. As a function of momentum transfer
the threshold term drops rather fast, whereas the linear terms falls off
much less. The quadratic term (with amplitude $d_2$) was negligible at
smaller momentum transfers and still quite small at values of -t $>$0.6 
GeV/c$^2$.  
As already discussed in former work (see e.g. \cite{Edel}), the 
background fits give rise to uncertainties in the absolute resonance 
cross sections up to 30 \%. However, the positions and widths
of the resonances are not much affected.  
Examples of the p-p and $\pi$-p fits at different momentum transfers are 
given in figs.~1 and 2.  

In fitting the spectra we used the $\Delta_{33}$(1232), the known N* 
resonances D$_{13}$(1520) and F$_{15}$(1680), and strong excitation 
of a resonance at 1400 MeV \cite{Andp,Edel,And}. Already in the former 
studies (see e.g.~\cite{Andp}) it has been questionned, 
whether this structure at 1400 MeV could be the Roper resonance P$_{11}$(1440) 
observed in $\pi$-N, because this peak is not correctly at the position of 
this resonance (masses of resonances were determined precisely in the 
spectrometer experiments \cite{Andp,Edel}). Other possibilities have been 
discussed (e.g.~special kinematic effects), which, however, are  
not very likely to explain the strongest resonance observed (which is  
stronger than $\Delta$(1232) excitation).
Remarkaby, at higher energies this resonance corresponds in mass and width 
closely to the Saturne resonance observed in $\alpha$-p scattering \cite{Mo92}
(see table~1).
Using all data above 9 GeV/c we obtain a centroid mass of the P$_{11}$ 
resonance of $1400\pm 10$ MeV with a width of $200\pm 20$ MeV (this is 
better determined than from $\alpha$-p, which has a strong form factor 
dependence). Further, the cross section of this resonance is peaked strongly 
at small momentum transfers, characteristic of L=0 excitation (see fig.~1). 

As discussed in sect.~1, evidence for two structures has been obtained 
\cite{MoZu} in the region of the lowest P$_{11}$ (Roper resonance), 
the breathing mode which is strongly excited in high energy p-p, and a second 
structure related to the $\Delta$ degree of freedom.
In the lower energy p-p data at about 6 GeV/c and $\pi$-p at 8 GeV/c (fig.~4)
we observe still a rather strong excitation of the $\Delta$(1232), which
decreases towards higher proton momenta. 
This is indicative of $\pi$-exchange, which should allow small  
excitation of the second structure. Indeed, in the spectra shown in fig.~4 
the P$_{11}$ peak is somewhat shifted to a higher mass (see table~1), 
which may be due to the effect of the second P$_{11}$ structure.
\begin{table}
\caption{Deduced resonance parameters of the P$_{11}$ resonance.} 
\begin{center}
\begin{tabular}{cl|ccc}
Reaction & beam momentum & ~ & $m_o$ &  $\Gamma$  \\

\hline
$\alpha$-p &\ \ 7 GeV/c     & ~ & 1390 $\pm$20 & 190 $\pm$30 \\
p-p        &\ \ 10-30 GeV/c & ~ & 1400 $\pm$10 & 200 $\pm$20 \\
$\pi$-p    &\ \ 16 GeV/c    & ~ & 1400 $\pm$15 & 200 $\pm$25 \\
\hline
p-p        &\ \ 6.2 GeV/c   & ~ & 1415 $\pm$10 & 205 $\pm$20 \\
$\pi$-p    &\ \ 8 GeV/c     & ~ & 1410 $\pm$15 & 200 $\pm$25 \\
\hline
$\pi$-N    &\ \ low E, ref.~\cite{phasepiN}& ~ & 1440 $\pm$30 & 360 $\pm$80 \\
$\gamma$-N &\ \ low E, ref.~\cite{gamma}   & ~ & 1463 $\pm$10 & 360 $\pm$20 \\
\end{tabular}
\end{center}
\end{table}

Results on the longitudinal e-p amplitude S$_{1/2}$ yield complementary
information on scalar N* excitations. Different from hadron scattering
longitudinal electron scattering becomes very small at higher energies.
In the inclusive p(e,e') data from SLAC~\cite{elec} the $\Delta$(1232) and N* 
resonances at 1520 and 1680 MeV have been observed, but the P$_{11}$
is missing. This can be taken as evidence, that the latter resonance has 
only a small transverse component at large momentum transfers. 

From recent exclusive e-p scattering experiments at JLab using a 
polarized electron beam, the longitudinal amplitude S$_{1/2}$ of the P$_{11}$ 
excitation could be extracted~\cite{elnew}. This amplitude is rather large, 
supporting our results of an important breathing mode excitation. 

{\bf 3. Decay of the P$_{11}$ at 1400 MeV from exclusive data}

Information on the decay of the strong P$_{11}$ resonance can be 
deduced also from high energy experiments. 
At a beam momentum of 6.6 GeV/c 4-prong events have been studied 
\cite{4prong} in p-p scattering, which are related to  
$p+p \rightarrow pp\ \pi^+ \pi^-$. In the invariant $\pi^+ \pi^-$ mass 
spectrum (ref.~\cite{4prong}, fig.~7a) a strong rise of the yield has been 
observed above the $2\pi$N threshold (see fig.~5), which is very different 
from phase space.
We calculated the $\pi^+ \pi^-$ mass spectrum corresponding to the 
resonances observed in figs.~1-4 and found, that indeed the resonance 
at 1400 MeV gives rise to a strong peak in the $\pi^+ \pi^-$ spectrum, 
whereas the resonances at higher mass are smeared out. This suggests
strong $2\pi$-N decay of this resonance. 

To calculate the complete 2$\pi$ spectrum, a background has 
been computed (dotted line) using a form consistent with that in eq.~(1) 
multiplied by the $2\pi$ phase space. As shown by the upper solid line 
in fig.~5 a reasonable description of the spectrum is obtained. Indeed, 
the strongest resonance contribution is due to the P$_{11}$, 
but the higher resonances D$_{13}$(1520) and F$_{15}$(1680) are needed also.
The latter contributions are in the order of 40 \% of that of 
the P$_{11}$ at 1400 MeV with a somewhat larger yield from the F$_{15}$(1680).
The D$_{13}$(1520) and F$_{15}$(1680) decay mainly into the $1\pi$N 
channel \cite{phasepiN} with a branching ratio of about 60 and 70 \%, 
respectively. These resonances are clearly observed in the 2-prong events 
of the p-p scattering data \cite{2prong}, whereas no clear signal of $1\pi$N
decay is observed for the resonance at 1400 MeV. 

In order to estimate the $2\pi$ decay branch of the P$_{11}$ resonance 
we have to take into account the fact, 
that the spectrum in fig.~5 is integrated over momentum transfer. 
In the low momentum transfer spectra in fig.~1 the ratio of P$_{11}$ to 
F$_{15}$ cross sections is about 3; integrating over all momentum 
transfers \footnote{for this integration we used the differential cross 
sections calculated in DWBA (see sect.~4)} up to 1.2 (GeV/c)$^2$ we 
obtain about equal yields (for larger values of -t the contributions are 
small). 
From this we extract a $2\pi$N branch of the P$_{11}$ of $75\pm 20$ \%. 
This is consistent with the result in ref.~\cite{MoZu} and that discussed 
above ($\sim $ 70 \%). Further, this supports former explanations of the 
1400 MeV resonance (see ref.~\cite{Andp}), being a highly inelastic 
resonance in $\pi$-N, which decays only weakly into the elastic channel.

{\bf 4. Description of the p-p differential cross sections }

To understand the differential cross sections we performed 
calculations within the framework of the {\it Distorted 
Wave Born Approximation} (DWBA) similar to those in ref.~\cite{MoSD}, 
assuming multi-gluon exchange of scalar structure as the predominant part 
of the effective interaction. First, elastic scattering was 
described in a double-folding potential similar to that in ref.~\cite{MoSD}. 
By fitting the experimental data, the strength of the effective interaction 
and the nucleon radius were determined. We obtained results, 
which represent an excellent 
extrapolation of the potential strengths and mean square radii 
deduced at lower energies (ref.~\cite{MoSD}, fig.~6 and 7). For a proton 
momentum of 15.1 GeV/c results are given in fig.~6 using volume 
integrals of the central real and imaginary potential I$_V$ and I$_W$ 
of about 520 and 640 MeVfm$^3$, respectively, and a nucleon mean square 
radius  $<r^2>_N$ of about 0.45 fm$^2$. 
At proton energies above 1 GeV the Lorentz boost gives rise to a 
contraction of the projectile in beam direction, leading to smaller values 
of $<r^2>_N$. On the other hand the systematics of p-p scattering at 
high energies shows the opposite effect, increasing slope parameters 
towards higher energies~\cite{Burq}, which may point to larger interaction 
radii. This may be understood by an increase of multiple scattering. 
Our results of $<r^2>_N$=0.45 fm$^2$ as compared 
to 0.66 fm$^2$ at smaller energies is well described by the boost effect,
indicating, that at the energies in question multiple scattering is still 
rather small. 

With these potentials inelastic (p,p') cross sections for resonance excitation 
were calculated in DWBA, the details of these calculations are similar to
those in ref.~\cite{MoSD}. As a result, only for L=0 transfer the t-dependence 
of the calculated cross section shows a diffraction pattern, with a slope
as oberved for the resonance at 1400 MeV. 
This is because the same partial waves contribute for the initial and final 
state wave functions. Thus, our calculations confirm quite model independently 
a P$_{11}$ assignment of this resonance. For larger L transfers the 
diffraction pattern is washed out 
due to the contribution of different partial waves in the overlap of 
initial and final state wave functions. This is observed for L=2 excitation 
of the F$_{15}$(1680), where the momentum transfer dependence is rather flat.
Here we want to mention, that the basic features of our DWBA approach are
quite consistent with the former theoretical approach in ref.~\cite{HT}, 
which, however, gave only a very qualitative account of the N* yields.

It is important to note, that the calculated t-dependence of the differential 
cross section is sensitive to details of the transition density. By using 
different forms we found, that the experimental cross sections in fig.~6 
can by described only by using a surface peaked transition density (quite
similar to that in ref.~\cite{MoSD}, fig.~10, dashed line). In sect.~5
an attempt is made to understand this transition density together with 
the data from e-p scattering. 
The absolute cross section of the breathing mode excitation is reproduced
with a monopole strength covering a significant fraction of the full energy 
weighted sum rule 
strength~\cite{MoSD,Mo}, in good agreement with the results of $\alpha$-p. 

{\bf 5. How can we understand the breathing mode excitation in p-p and
  e-p?}

The dominance of multi-gluon exchange~\cite{MoSD} in p-p scattering at 
the energies in question can be taken as indication, that this reaction 
probes to a large extent the multi-gluon structure of the nucleon 
(in addition to valence quark excitation). 
Support for strong multi-gluon contributions to the breathing mode 
excitation is also obtained from the deduced hadron 
compressibilities~\cite{Mo}, which were found to be flavor and isospin 
independent, and thus quite likely associated with the multi-gluon 
field of hadrons. Such gluonic contributions are also expected from the basic
structure of QCD, which involves strong self-interactions of gluons.
 
In the breathing mode excitation a large fraction of the energy weighted sum
rule strength is observed; this allows to approximate the transition 
density $\Delta\rho_o$(r) in a fluid-dynamical picture by dynamical 
distortions of the g.s.~density $\rho$(r) with respect to the central
density $\rho_o$ and the density fall-off $m$.
Using for this a form $\rho(r)=\rho_o exp(-m r^\kappa)$ 
(with values of $\kappa$ between 1 and 2) gives
\begin{equation}
\Delta\rho_o(\vec r)=[\delta \rho_o \frac{\partial\rho(r)}{\partial \rho_o} 
f_o(r)+\delta m \frac{\partial\rho(r)}{\partial m}\ +(h.o.)]\ 
Y_o(\theta,\phi)\ ,
\end{equation}
The amplitudes $\delta \rho_o$ and $\delta m$ were varied to obtain a 
reasonable fit of the differential cross sections, further $f_o(r)$=1 was
used. 

A good description of the p-p differential cross sections in fig.~6 is 
obtained asssuming only a distortion $\delta m$$\neq$0 of the surface term. 
This is very different from our expectation for a radial excitation 
mode (see~\cite{MoSD}) with the constraint $\int \Delta\rho_o(r)\ r^2 dr=0$,
for which $\delta \rho_o$ and $\delta m$ $\neq$0. To be able to compare
the extracted transition density with e-p scattering, boost effects should 
be removed. For this, the resulting transition density was revaluated for 
a density yielding $<r^2>_N$=0.66 fm$^2$, given by the dot-dashed line 
in the upper part of fig.~7.

As pointed out, the longitudinal e-p amplitude $S_{1/2}$~\cite{elnew} 
yields complementary information on the breathing mode excitation.
The charge transition density was assumed of the same form as eq.~(2),
using a charge density consistent with ref.~\cite{Kelly} and the 
additional constraint $\int \Delta\rho_{ch}(r)\ r^2 dr=0$ 
(charge conservation), which requires $\delta \rho_o$ and $\delta m$ $\neq$0. 
A fit of the experimental data is given in the lower part of 
fig.~7 with a transition density as given by the solid line in the upper part. 
The transformation of the transition density to a Q$^2$ dependence was
made according to ref.~\cite{Kelly} (the two lines are for 
$\lambda$=1 and 2 and may indicate the size of uncertainties in the 
relativistic transformation). Because for the valence quarks ($q^3$) 
also $\int \Delta\rho_{q^3}(r)\ r^2 dr=0$, the charge transition density 
corresponds essentially to that of the valence quarks. Our results are 
consistent with lattice QCD calculations~\cite{latt} indicating valence 
quark excitation in the low lying P$_{11}$ resonance. Actually, from 
recent lattice calculations~\cite{lattwf}, information on the radial wave 
functions of the nucleon ground and first excited state could be extracted. 
These calculations show a node in the P$_{11}$ wave function, supporting 
the breathing mode picture.

The deduced transition density of the valence quarks, 
$\Delta\rho_{q^3}(r) \approx \Delta\rho_{ch}(r)$ (solid line in fig.~7, 
upper part) is very different from the matter (multi-gluon plus $q^3$) 
transition density derived from p-p (dot-dashed line).
This suggests strongly, that in p-p the relative $q^3$ coupling $R_{q^3}$ 
(with respect to the coupling to the multi-gluon field)
is rather small. Assuming for $R_{q^3}$ a value of 0.3, the multi-gluon 
(or sea quark) part of the transition density is given by the dashed line. 
This component is almost an order of magnitude stronger than the valence
quark component in p-p, and with opposite phase. Further, it shows also 
``breathing'' (node in the transition density), indicating that strong 
sea quark components due to the multi-gluon field are already present in 
the g.s.~density (again confirming our assumption of a strong multi-gluon 
field). Qualitatively similar results are obtained for even smaller 
values of $R_{q^3}$ (which are probably more realistic). Further, 
the surface peaked matter transition density (dot-dashed line) indicates 
in addition creation of multi-glue (or $q^{2n}\bar q^{2n}$) components 
out of the g.s.~vacuum. 

{\bf 6. Summary}

The present analysis of high energy p-p scattering 
data has confirmed our evidence for a ``scalar'' P$_{11}$ excitation, 
which covers about the full energy weighted sum rule (breathing mode). 
The properties of this remarkable excitation ({\it L=0 without spin-isospin 
flip, mass $\sim$1400 MeV, width $\sim$200 MeV, dominent 2$\pi$-N decay}) 
are different from those of the 
Roper resonance observed in low energy $\pi$-N and $\gamma$-N 
({\it dominant L=0 spin-isospin (M1) excitation, mass $\sim$1440 MeV, 
width about 360 MeV, dominent 1$\pi$-N decay}); this supports strongly 
the two resonance picture of the lowest P$_{11}$ (Roper) resonance.
In the low energy reactions $\pi$-N and 
$\gamma$-N the breathing mode has not been detected. For $\pi$-N this can be
understood by the fact, that the S-wave interaction is strongly reduced 
as compared to P-wave excitation; in addition, the coupling to 
$\pi$-N is further suppressed due to the dominant 2$\pi$N 
branch of this resonance. In the case of $\gamma$-induced reactions a pure 
scalar L=0 state cannot be excited. However, both in $\pi$-N and
$\gamma$-N strong coupling to the $\Delta$ degree of freedom
is observed, which allows excitation of the second P$_{11}$ structure 
understood as $\Delta$ excitation of the $\Delta$(1232).

The evidence for the breathing mode is further supported by recent polarized
e-p scattering data, from which the longitudinal amplitude $S_{1/2}$ 
has been extracted. An attempt has been made to understand the data on p-p 
and e-p consistently. For this it was necessary to assume the existence 
of strong multi-gluon contributions in the nucleon density consistent with the 
vacuum structure of Yang-Mills theory. 
In longitudinal electron scattering only charge contributions (mainly 
valence quarks) are seen, whereas the interaction in p-p probes mainly the
multi-gluon structure (giving rise to strong sea quark contributions). This 
yields evidence for a rather complex multi-quark structure of baryons. 

Future experimental work should provide more details on the interplay of quark
and gluonic structure. This is possible in detailed studies of selective 
hadronic and electromagnetic reactions, as discussed in the present paper. 
Therefore, complementary to the large efforts made at Jlab
with polarized electron scattering experiments, selective N* excitation 
in p-$\alpha$ scattering 
with clear separation of the important 2$\pi$N decay channels, 2$\pi(s)$-N, 
2$\pi(p)$-N, and $\pi$-$\Delta$, and other channels are planned at COSY.

{\bf Acknowledgement}

We thank Piotr Decowski for many comments and suggestions.

\begin{figure} [ht]
\centering
\includegraphics [height=18cm,angle=0] {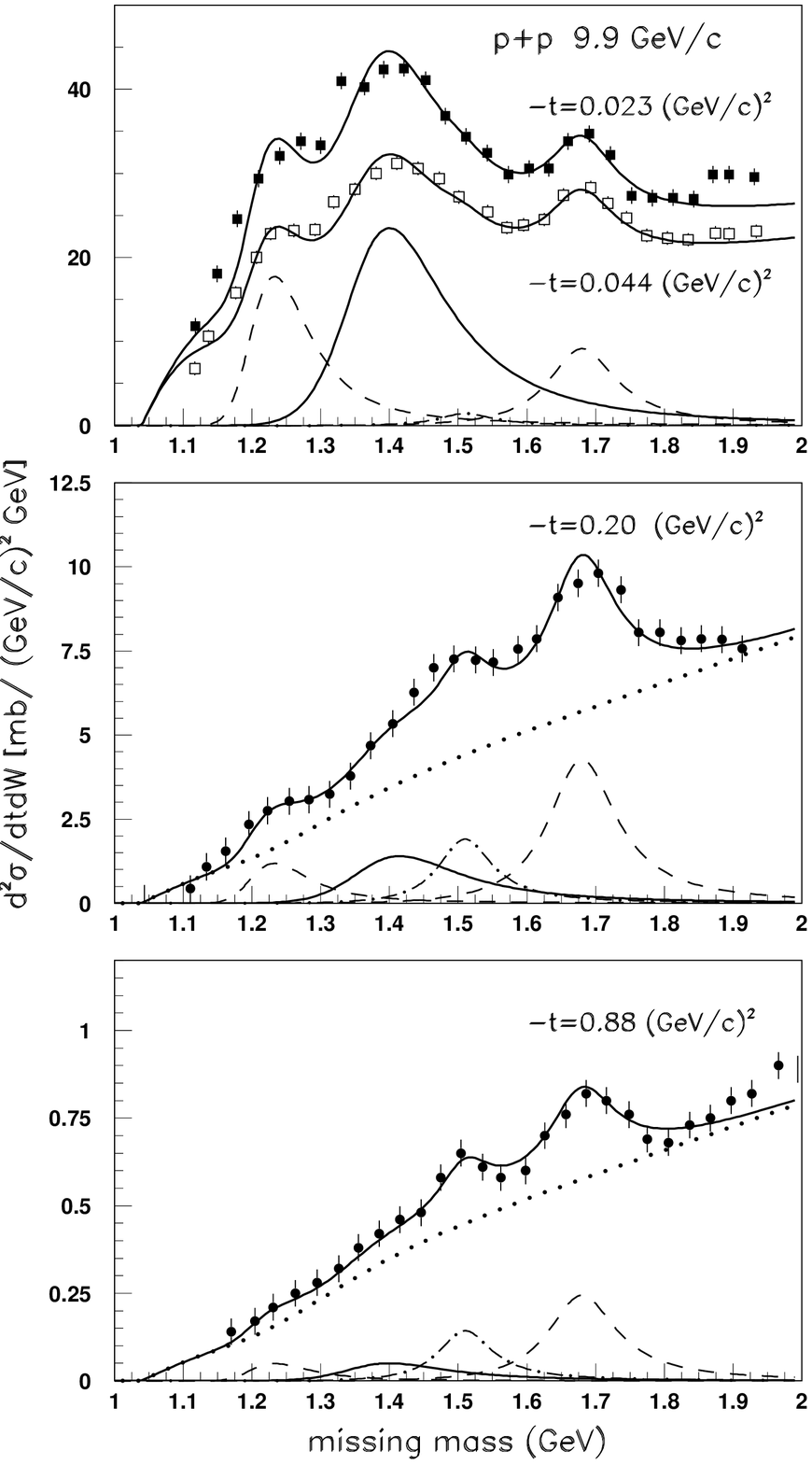}
\label{fig1}
\caption{Missing mass spectra for p-p$\rightarrow$p'+x at a beam momentum of 
9.9 GeV/c from ref.~\cite{Edel} in comparison with 
resonance fits (solid lines) using a background shape (dotted lines)  
given by eq.~(1). The separate resonances are given, in particular strong
excitation of the P$_{11}$ at 1400 MeV at small momentum transfer is 
indicated by solid lines.}
\end{figure}

\begin{figure} [ht]
\centering
\includegraphics [height=16cm,angle=0] {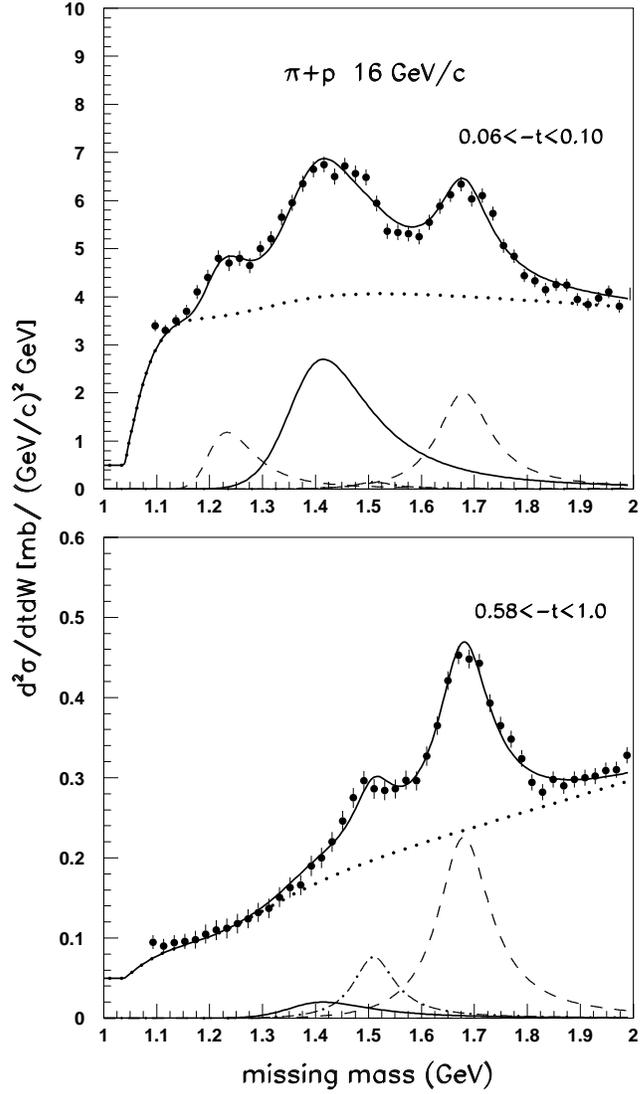}
\label{fig2}
\caption{Missing mass spectra for $\pi-p\rightarrow\pi'+x$ at a beam 
momentum of 16 GeV/c from ref.~\cite{Edel} in comparison with 
resonance and background fits similar to those in fig.~1.}
\end{figure}

\begin{figure} [ht]
\centering
\includegraphics [height=18cm,angle=0] {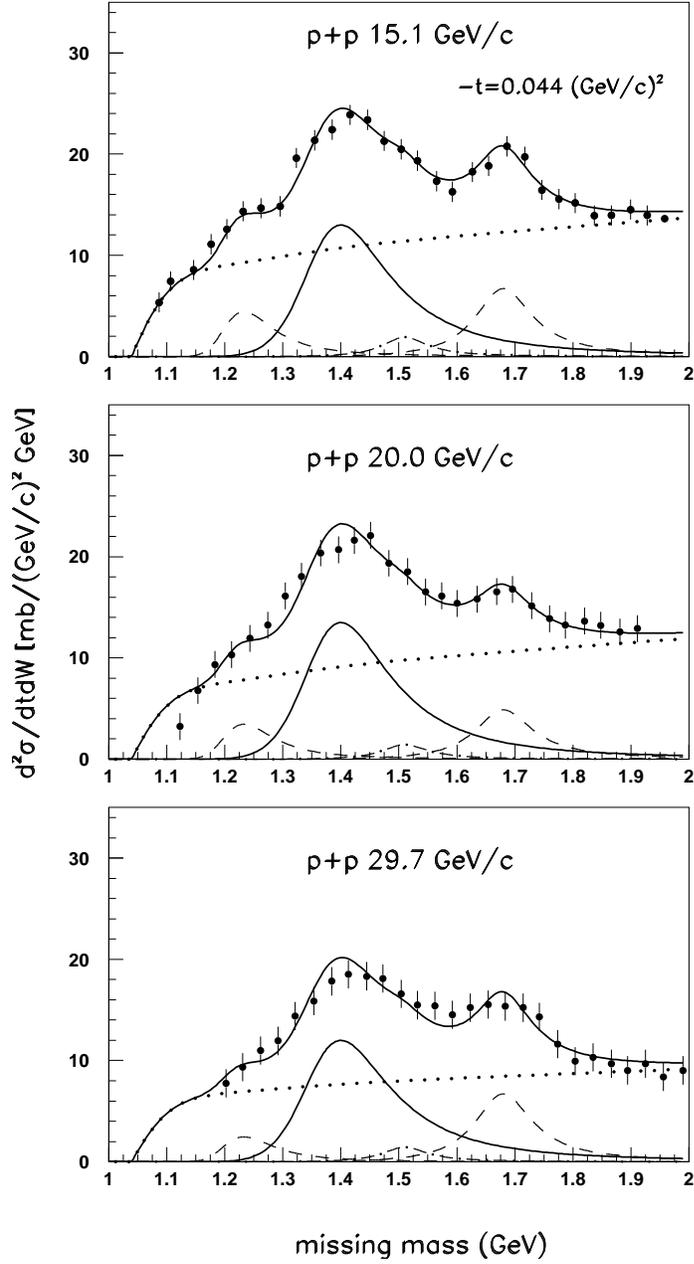}
\label{fig3}
\caption{Missing mass spectra for p-p$\rightarrow$p'+x at beam 
momenta of 15.1, 20.0, and 29.7 GeV/c at a momentum transfer of 0.044
(GeV/c)$^2$ from ref.~\cite{Edel} in comparison with 
resonance and background fits similar to those in fig.~1.}
\end{figure}

\begin{figure} [ht]
\centering
\includegraphics [height=16cm,angle=0] {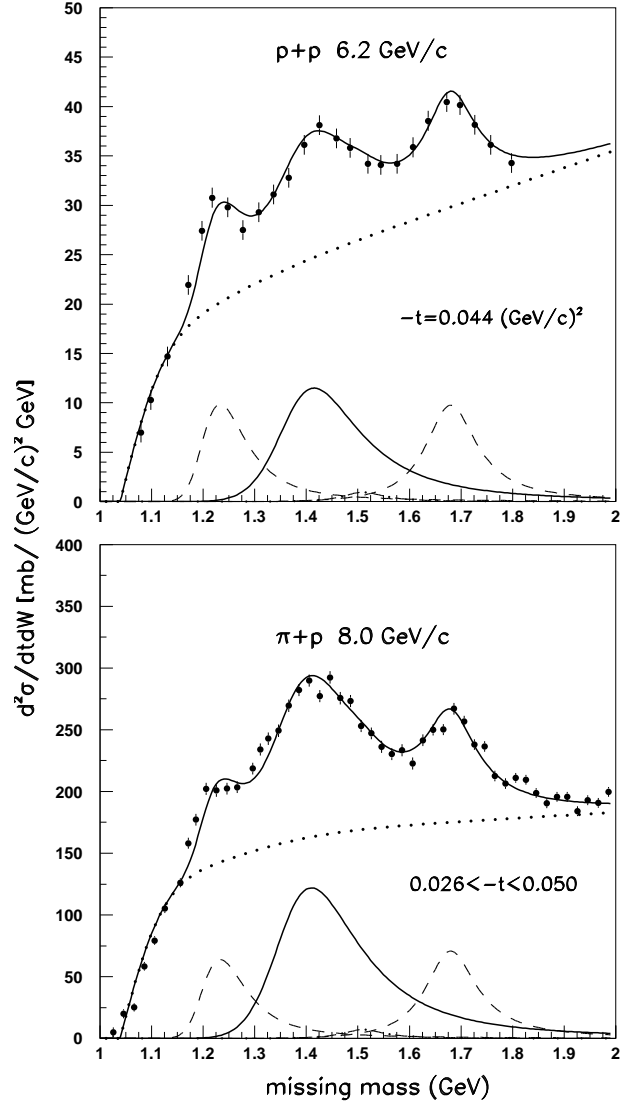}
\label{fig4}
\caption{Missing mass spectra for p-p$\rightarrow$p'+x at a beam momentum of
6.2 GeV/c (upper part) and $\pi-p\rightarrow\pi'+x$ at a beam 
momentum of 8.0 GeV/c from ref.~\cite{Edel,And} in comparison with 
resonance and background fits similar to those in fig.~1.}
\end{figure}

\begin{figure} [ht]
\centering
\includegraphics [height=17cm,angle=0] {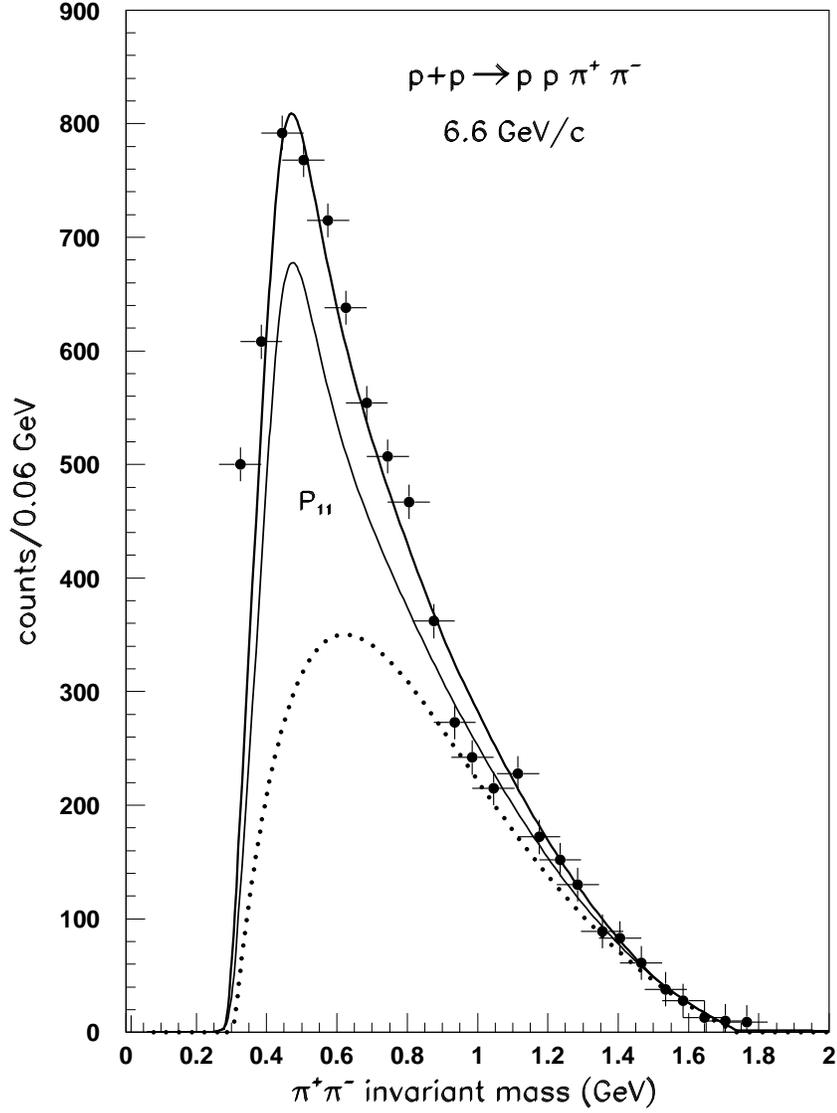}
\label{fig5}
\caption{Invariant $\pi^+ \pi^-$ mass spectrum from 
the data in ref.~\cite{4prong} in comparison with a resonance fit (upper
solid line) consistent with fig.~1. The 2$\pi$ background without 
and with the contribution of the P$_{11}$ at 1400 MeV is given by the 
dotted and lower solid line, respectively.}
\end{figure}

\begin{figure} [ht]
\centering
\includegraphics [height=18cm,angle=0] {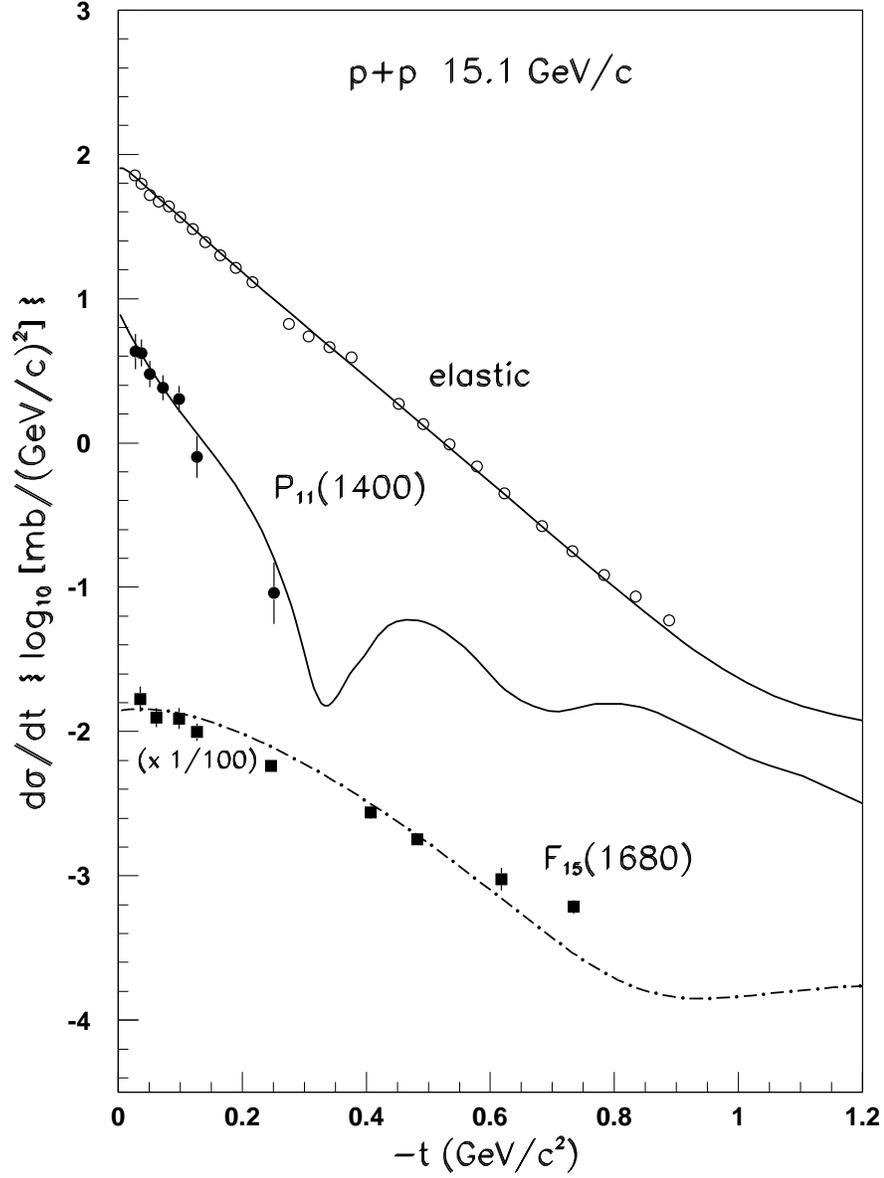}
\label{fig6}
\caption{Differential cross sections for elastic scattering and inelastic
excitation of resonances at 1400 and 1680 MeV for 15.1 GeV/c p-p scattering
from ref.~\cite{Edel} together with the result of our DWBA calculations.}
\end{figure}

\begin{figure} [ht]
\centering
\includegraphics [height=17cm,angle=0] {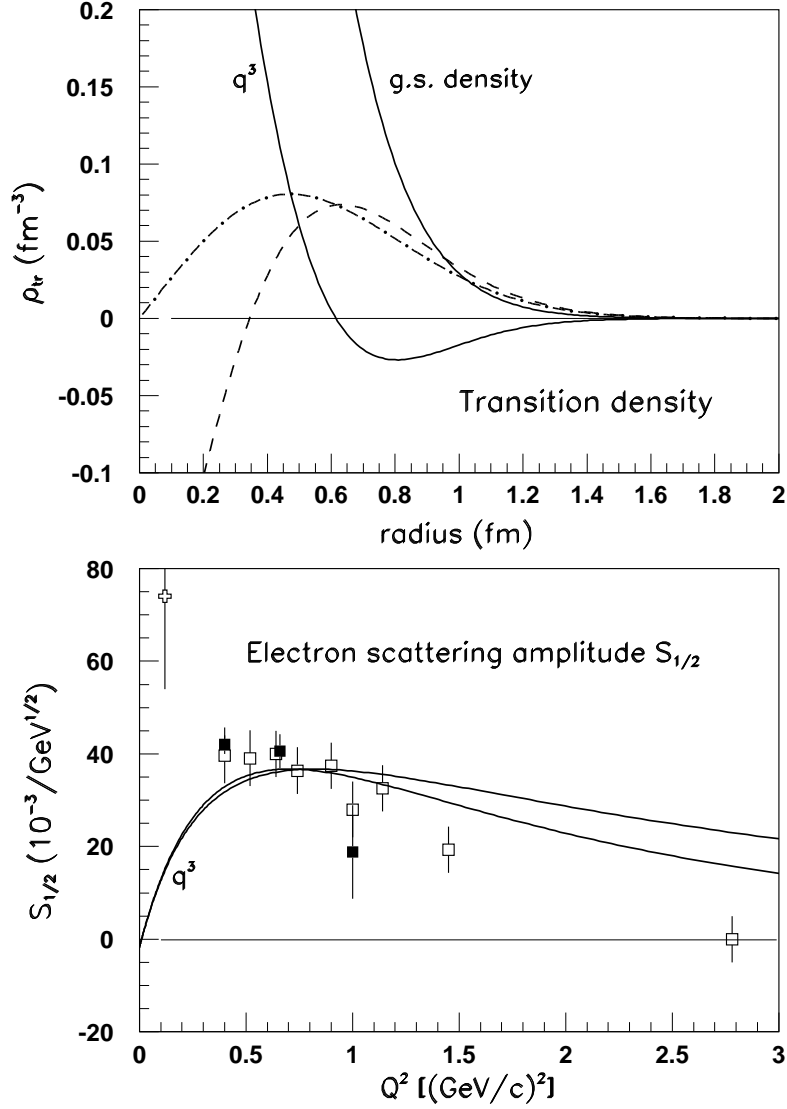}
\label{fig7}
\caption{Transition densities derived from p-p and e-p data (upper part)
and longitudinal e-p scattering amplitude $S_{1/2}$ with the data from 
ref.~\cite{elnew} (lower part). The solid lines (apart from the g.s.~density)
describe $S_{1/2}$, whereas the dot-dashed line results from a fit of p-p.
The dashed line indicates the sea quark contributions arising from 
the multi-gluon field. }
\end{figure}

\end{document}